\documentstyle[epsf,aps,preprint]{revtex}
\begin{document}
\widetext
\draft
\tighten

\title{Bound $q^2\bar q^2$ states in a
constituent quark model\thanks{Submitted to Z. Phys. A} }

\author{M. W. Beinker\thanks{E-Mail: beinker@ptprs10.phy.tu-dresden.de}}
\address{Institut f\"ur Theoretische Physik, TU
Dresden}

\author{ B. C. Metsch and H. R. Petry}
\address
{Institut f\"ur Theoretische Kernphysik,
Universit\"at Bonn}

\maketitle

\begin{abstract}
We consider the existence of bound systems
consisting of two quarks and two antiquarks
($q^2\bar q^2$) within the framework of a
constituent quark model. The underlying quark
dynamics is described by a linear confinement potential and
an effective $q^2\bar q^2$ interaction which has its
origin in instanton effects of QCD. We calculate the
spectra and examine the internal structure of the
states found.
\end{abstract}

\section{Introduction}
Most hadrons found experimentally have been
identified as mesons ($q\bar q$) or baryons ($q^3$)
and their masses and
decays can be described fairly well in
various quark models.  However, there are
still a few so-called exotic mesons which can not be
interpreted as $q\bar q$ states. For the lightest
exotic mesons $f_0(975)$ and $a_0(980)$ an
underlying $q^2\bar q^2$ structure (or $K\bar K$
molecule), besides other interpretations
(e.g. \cite{gribov}), has been discussed
\cite{jaffe,isgur1,isgur2,oka}, but could not been proved
yet unambiguously. Recent experiments and data has renewed
the interest in these exotic states
\cite{speth,zubov,kerbikov,wyech}.

We will discuss the existence of bound $q^2\bar
q^2$ states within a non-relativistic constituent
quark model, which was successfully used to describe
meson and baryon mass spectra \cite{blask} and
decays. We can compare the states of this four particle
problem with uncorrelated two-meson states and thus obtain a
first clue whether the calculated states
in reality would be bound or dissociated. In
addition, the model is just simple enough to make
the calculation of this real four particle problem
possible.

We consider the three lightest quark flavours
$u$, $d$ and $s$, where the constituent
quark masses of $n$ and $s$ ($n$ stands for a
$u$ or $d$) are assumed to differ.
The potential part of the Hamiltonian consists of a
sum of two-body forces. This contains a long range,
colourdependent confinement part and a short range
instanton induced interaction, which is generalized to three
flavours \cite{blask}.

The hamiltoninan was calculated in a full basis of
solutions of the $3\times3$-dimensional harmonic
oscillator. Solving the eigenvalue problem and
by a variational principle yield the mass
spectra and the corresponding eigenstates for further
examination of the internal structure of the states.

\section{The Model}
For calculating the $q^2\bar q^2$ states, we used a
Hamiltonian of the form
\begin{eqnarray}
\label{hamil}
H=M+K+V_{\rm conf}+W
\end{eqnarray}
where
\[
M=\sum_i m_i
\]
is the sum of the four constituent quark masses and
\[
K=\sum_i\frac{p^2_i}{2m_i}-\frac{P^2}{2M}
\]
the kinetic energy for the relative motion.

As confinement potential $V_{\rm conf}$, we choosed
a linear rising potential with the relative
quark-(anti)quark distance:
\[
V_{\rm conf}(r_{ij})=\sum_{i<j}F_C(a_{ij}+b_{ij}r_{ij}).
\]
The offset $a_{ij}$ and the slope $b_{ij}$ are different for
$qq$ and $q\bar q$ confinement. As noted in \cite{grom}
$a_{q\bar q}$ and $b_{q\bar q}$ are not independent. The
dependence of $b_{qq}$ from $b_{q\bar q}$ is given by
geometrical reasons~\cite{volker}.  The following relations
should hold:
\begin{eqnarray*}
a_{q\bar q} & = & -2\sqrt{b_{q\bar q}} \\
b_{qq} & = & 0.5493^{-1}b_{q\bar q}
\end{eqnarray*}
Although the parameters were independently
varied, the above relations in the present approach
are fullfilled with
great accuracy.  They are given in
table~\ref{para}. We consider two different kind of colour
dependencies $F_C$ of $V_{\rm conf}$, which in the meson
or baryon case are identical.  In the first case, $F_C$ is
identical to the colour singlett projector $P_C^1$
\cite{isgur2,oka}:
\[
V_{\rm conf}^1(r_{ij})=
\sum_{i<j}P_C^1(a_{ij}+b_{ij}r_{ij})
\]
whereas in the other case we have:
\[
V_{\rm conf}^\lambda(r_{ij})=\sum_{i<j}
\frac{\lambda_i}{2}\cdot
\frac{\lambda_j}{2}
(a_{ij}+b_{ij}r_{ij})
\]
with $\lambda_i$
is a Gell-Man matrix.
The $V^1_{\rm conf}$ potential acts only on colour
singletts. In the case of $q^2\bar q^2$ only the
four $q\bar q$ pairs will be affected.  Since it is
an purely attractive potential
two distinct $q\bar q$ colour singlett
states (which we call quasi-mesons) will always have
a net attraction, because also $q\bar q$
pairs with $q$ and $\bar q$ from different
quasi-mesons have always colour singlett
contributions yielding unwanted long range forces.

The $V^\lambda_{\rm conf}$ potential has a better
behaviour for $R\rightarrow\infty$, $R$ is the
distance of the two quasi-mesons. As was shown
\cite{gavela}, it goes with $R^{-3}$ as $R$ grows
large which is a consequence of the
repulsive part of $V^\lambda_{\rm conf}$, which is
an effect without experimental evidence.
In addition
$V^\lambda_{\rm conf}$ allows a better comparison
with earlier results of other authors
\cite{isgur2,oka}.

Finally,
\[
W=H_{q_1q_2}+H_{\bar q_3\bar q_4}+
H_{q_1\bar q_3}+H_{q_2\bar q_4}
+H_{q_1\bar q_4}+H_{q_2\bar q_4}
\]
is the residual quark-(anti)quark interaction. The
$q_1q_2$ term is of the form
\[
H_{q_1q_2}=-\tilde g(P^{S=1}P^C_6+2P^{S=0}P^C_{\bar 3})\delta^3(\vec r)
\]
where the flavour matrix $\tilde g$ is shown in table~\ref{gqq}, $P^S$
are spin projectors on the relative $qq$ spin state
and $P^C$ denotes a colour projector. $\vec r$ is
the relative distance of the two quarks.  The $\bar
q\bar q$ term is of the same form, if one changes
all colour to anticolours and all flavour to
their antiflavour indices.  All $q\bar q$ terms are of
the form
\[
H_{q\bar q}=
\hat g\left(\frac32P^{S=1}P^C_8+P^{S=0}
\left(\frac12P^C_8+8P^C_1\right)\right)
\delta^3(\vec r)
\]
where the flavour depending $\hat g$ is given in table~\ref{gqQ} (for the
calculation of this residual interaction from an
effective Lagrangian see \cite{blask}).

The matrices $\tilde g$ and
$\hat g$ include the total flavour dependence of
$W$. As one can see from table~\ref{gqq},
$\tilde g$ contains a projector on flavour antisymmetric
states, while $\hat g$ generates a flavour mixing
for $T=0$ states, which allows to describe the
$\eta$-$\eta'$ splitting in the meson case, and
lowers the energy of the isovector state.

The more complicated structure of $W$ for
$q^2\bar q^2$ in comparison to the meson or baryon
case is a consequence of the fact that there exist
two distinct colour singletts for $q^2\bar q^2$ and
consequently the terms including $P^C_6$ or $P^C_8$ do not
vanish.
The residual interaction is a pure contact force and
as such leads to an unbound Hamiltonian. Therefore, we
have chosen to regularize the interaction by
replacing the $\delta$ distribution by a Gaussian
\[
\delta^3(\vec r)\rightarrow\frac1{\Lambda^3}
\frac1{\pi^{\frac32}}
e^{-\frac{r^2}{\Lambda^2}}\delta_{L,0}
\]
where $\Lambda$ can be interpreted as an effective
range of the pairing force. We note that a
derivation of the 't Hooft force going beyond the one
loop approximation which was used, is expected to
such a finite
range~\cite{blask}.

For the calculation of the $q^2\bar q^2$ mass spectra
and the eigenstates, we proceed as
follows: The matrix elements of the
Hamiltonian~(\ref{hamil}) are calculated within a
spin-flavour $SU(6)$, $O(3)$ oscillator basis for
each relative coordinate comprising $N=8$ oscillator
excitations in total. We should note that this size of the
model space is still too small for reaching convergence of
the spectra.

Calculations with $N>8$ oscillator excitation
were limited by the calculational effort
However, the
comparison with meson calculations shows
that the lowest eigenvalues are calculated with an
accuracy of a few MeV.

We obtained the $q^2\bar q^2$ spectra by application
of the variational principle, i.e. diagonalisation
of the Hamiltonian in the oscillator basis and
minimization of the lowest energy eigenvalue with
respect to the oscillator length parameter. This is
consistent to the earlier meson and hadron
calculations in~\cite{blask}.  The parameters
entering the present calculation are given in
table~\ref{para}. These parameter were fitted to the
meson and baryon spectra.

\section{The Spectra}

\begin{figure}
\centering
\leavevmode
\epsfysize=0.8\textheight
\addtolength{\epsfysize}{-1cm}
\epsffile{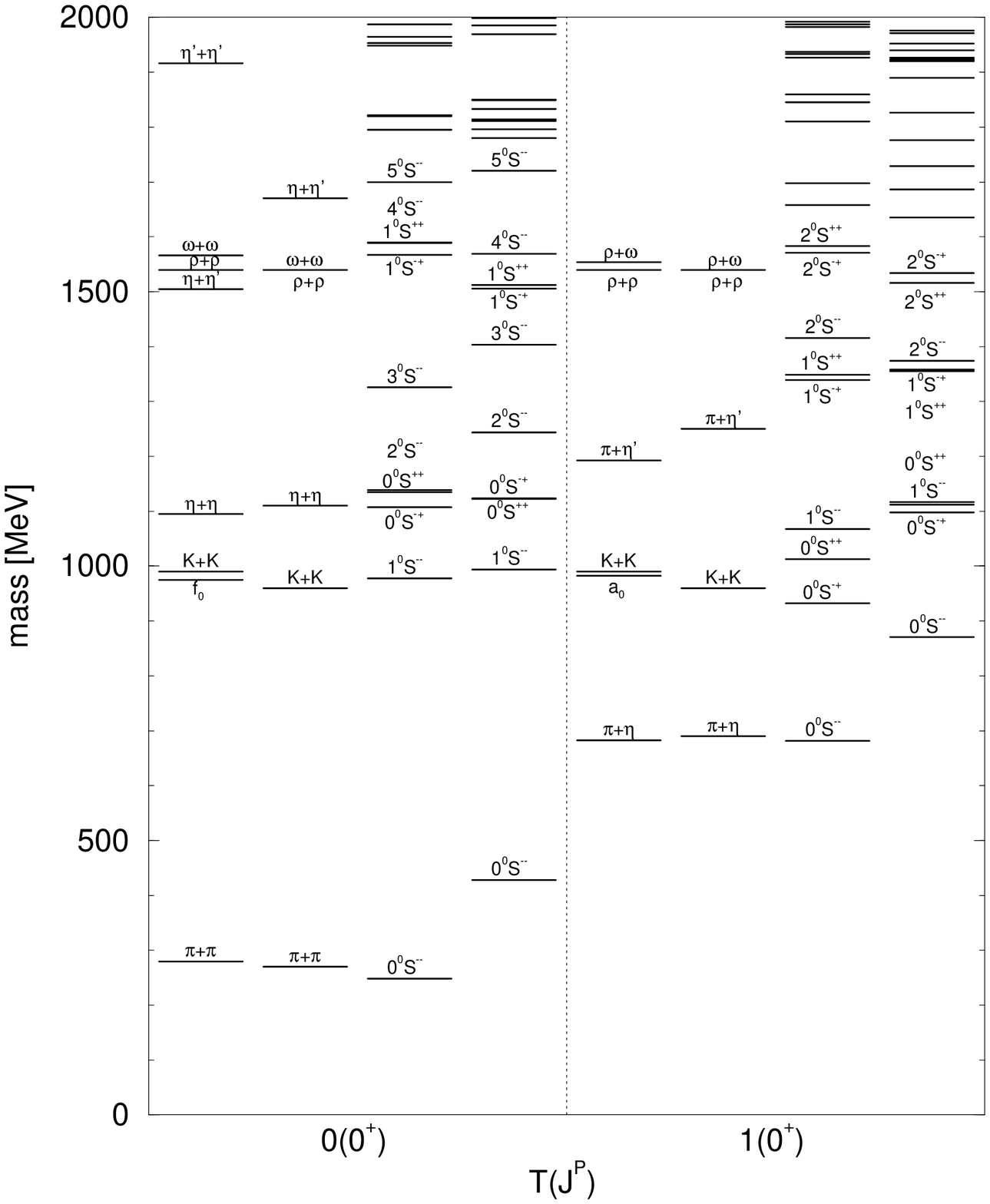}
\caption{
Spectra for scalar-isoscalar and
vector-scalar $q^2\bar q^2$-systems.
The subcolumns show from left to right
the sum of the experimental masses of two
mesons, the sum of two meson masses calculated
with the the same model
in the single meson sector,
the calculated $q^2\bar q^2$ spectra with
 $V^1_{\rm conf}$ and, finally,
the spectra calculated with
$V^\lambda_{\rm conf}$
\label{spec}}
\end{figure}

Fig.~\ref{spec} shows the mass spectra in the most
interesting scalar isoscalar and isovector
channel. The columns for each channel are devided
into four subcolumns, which show from left to right
the sum of the experimental masses of two
mesons, the sum of two meson masses calculated
with the the same model
in the meson sector,
the calculated $q^2\bar q^2$ spectra with
 $V^1_{\rm conf}$ and, finally,
the spectra calculated with
$V^\lambda_{\rm conf}$.  In addition, the very left
subcolumns shows the masses of $f_0(975)$ and
$a_0(980)$, the most popular $q^2\bar q^2$
candidates.

The $q^2\bar q^2$ eigenvalues are denoted by
$n^{2S+1}L^{p_{12}p_{34}}$ where $S$ is the total
spin, $L$ the total angular momentum and $p_{12}$
resp. $p_{34}$ the exchange symmetrie of $q_1$ with
$q_2$ and  $\bar q_3$ with $\bar q_4$ respectively.
$n$ counts
all states with identical $S$, $L$, $p_{12}$ and
$p_{34}$.

The two meson-state masses, which are compared with the
$q^2\bar q^2$ spectra, are choosen to have the same
observable quantum numbers as the $q^2\bar q^2$
states with the two mesons in a relative
$s$-wave.  A main diffiulty of our $q^2\bar q^2$
model is, that we can not describe two free mesons
in a finite model space with a finite number of
oscillator excitations.
Otherwise, it would be impossible to
calculate the Hamiltonian. It is therefore
neccessary to examine the internal structure of the
$q^2\bar q^2$ states in detail
to differentiate real bound $q^2\bar
q^2$ states from quasi free two meson states.

As one can see from fig.~\ref{spec}, the $q^2\bar
q^2$ spectra describe in good approximation the
lowest two meson mass sum for the
spectra calculated with the $V^1_{\rm conf}$
potential. The lowest state $0^0S^{--}$ lies under
the $m_\pi+m_\pi$ treshold which is due to the
strong long range meson-meson attraction
from the $V^1_{\rm conf}$ potential.
Since we consider this long range attraction to be
unphysical, this does not
imply that $0^0S^{--}$ is in reality a bound $q^2\bar q^2$
state. The eigenstates calculated with the
$V^\lambda_{\rm conf}$, especially the lowest one,
have higher energies due to the repulsive part of
this potential.

As a consequence of the discretization of the two
meson continuum in our variational approach,
both spectra contain more
eigenstates in the energy region shown than the
sum of the masses of two mesons
(with the two mesons in a $s$-wave)
would
predict. In other words, the spectra
alone do not allow to draw any conclusions wether
there exist any ``real'' $q^2\bar q^2$ bound states
within this model or not.  A closer examination of
the internal structure of the calculated eigenstates
is therefore neccessary.

\section{Internal Structure}

\begin{figure}
\centering
\leavevmode
\epsfysize=0.45\textheight
\epsffile{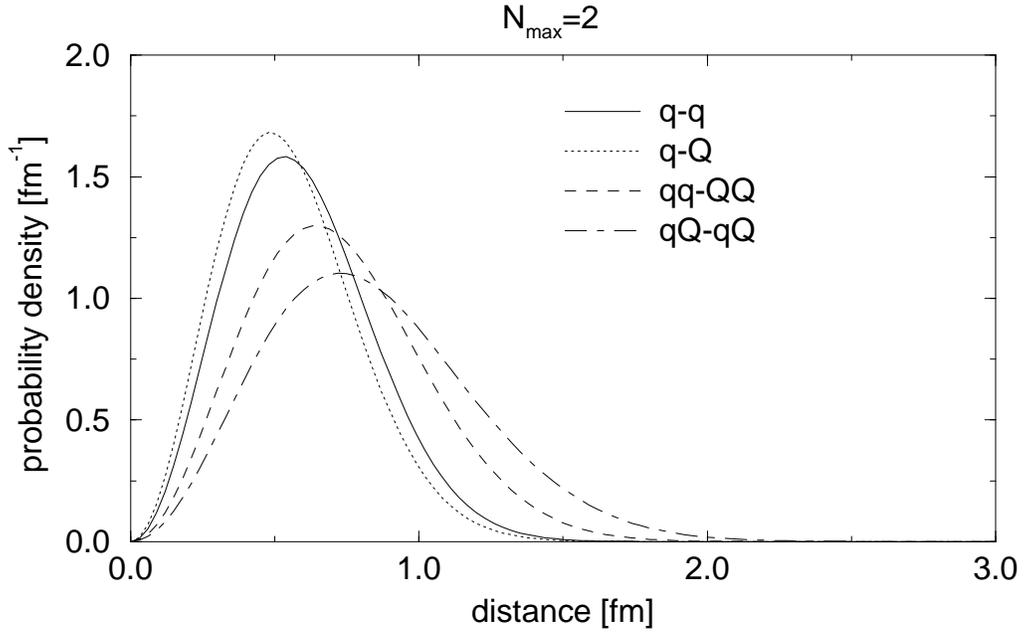}\\
\leavevmode
\epsfysize=0.45\textheight
\epsffile{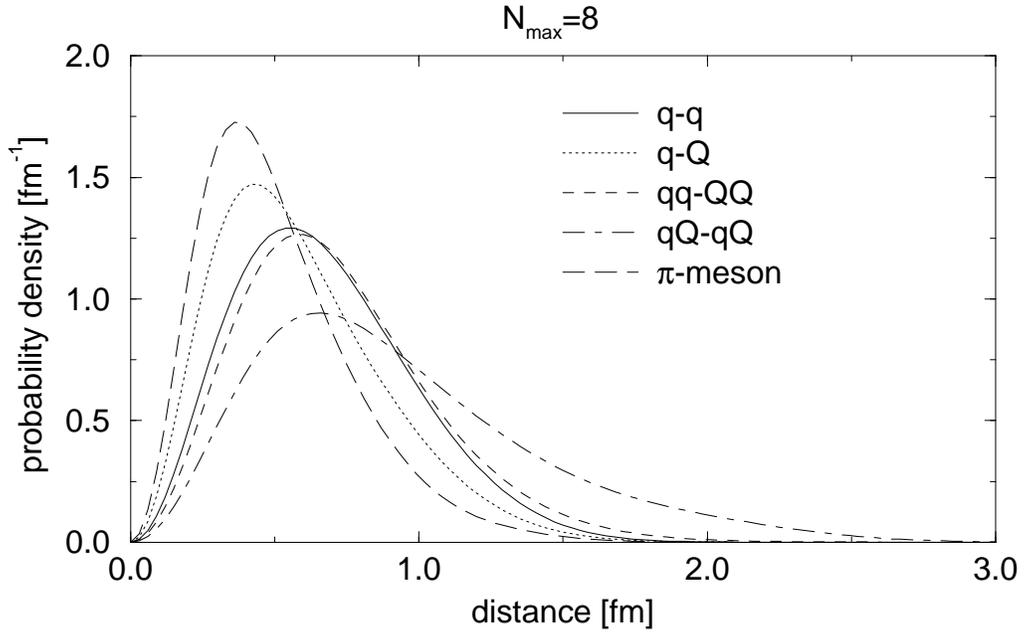}
\caption{
Density distributions in the channel $0(S^+)$
in dependence of $N_{max}$.
The distributions belong to the ground state. A
{\sf Q} denotes here an anti-quark $\bar q$.
For comparison the $q$-$\bar q$ distribution
of the $\pi$-meson calculated in the same
framework is shown.
\label{densnmax}}
\end{figure}

To obtain such a closer view of the interpretation of
the internal structure
of the calculated $q^2\bar q^2$ states, we calculated various
probability densities. Let us consider the lowest
eigenstate $0^0S^{--}$. (Unless stated otherwise we
refer to the calculation with the $V^1_{\rm conf}$
potential.)  Fig.~\ref{densnmax} shows the densities
for the various relative distances with
$N_{max}=2$ and $8$.
These densities were calculated by
summing over
all internal quantum numbers and integrating the
$q^2\bar q^2$ probability density
$\Psi_\alpha^\dagger(\vec u_1,\vec u_2,\vec u_3)
\Psi_\alpha(\vec u_1,\vec u_2,\vec u_3)$
over all relative coordinates except the relative distance
under investigation, with $\vec u_i$ denoting a
relative coordinate
(e.g. the $q$-$\bar q$-distance) and $\alpha$ a
multiindex for all internal quantum numbers.

The $q\bar q$ distribution is calculated as:
\begin{eqnarray}
\rho_{q\bar q}(\vec u)=\sum_\alpha
\int d^3\vec u_{q_2,\bar q_4}d^3 \vec u_{q_1\bar q_3,q_2\bar q_4}
\left|\Psi_\alpha(\vec u,\vec u_{q_2,\bar q_4},
 \vec u_{q_1\bar q_3,q_2\bar q_4})\right|^2,
\end{eqnarray}
since all four pairs have the same
distribution.
It is observed that all
densities in fig.~\ref{densnmax} are spreading with
growing $N_{max}$. While the $q$-$\bar q$ distance
(or inner quasi-meson radius) distribution keeps its
form, the $q\bar q$-$q\bar q$ (or quasi-meson)
distance distribution is mostly spread and its
form seems to go over in a uniform distribution with
rising $N_{max}$, only slightly modified by
the long range part of the $V^1_{\rm conf}$ potential, as
mentioned above.

For higher excited states the $q\bar q$-$q\bar q$
distribution gets more complex and broader while
the $q$-$\bar q$ distribution again ist mostly localised,
and the same conclusions as above can be
drawn.
Comparison with the distribution
of $\pi$-meson radius within the same model
shows that it has nearly the same form as the
$q$-$\bar q$ distribution of the lowest $q^2\bar
q^2$ state $0^0S^{--}$
(see fig.~\ref{densnmax}).

\begin{figure}
\centering
\leavevmode
\epsfxsize=\textwidth
\epsffile{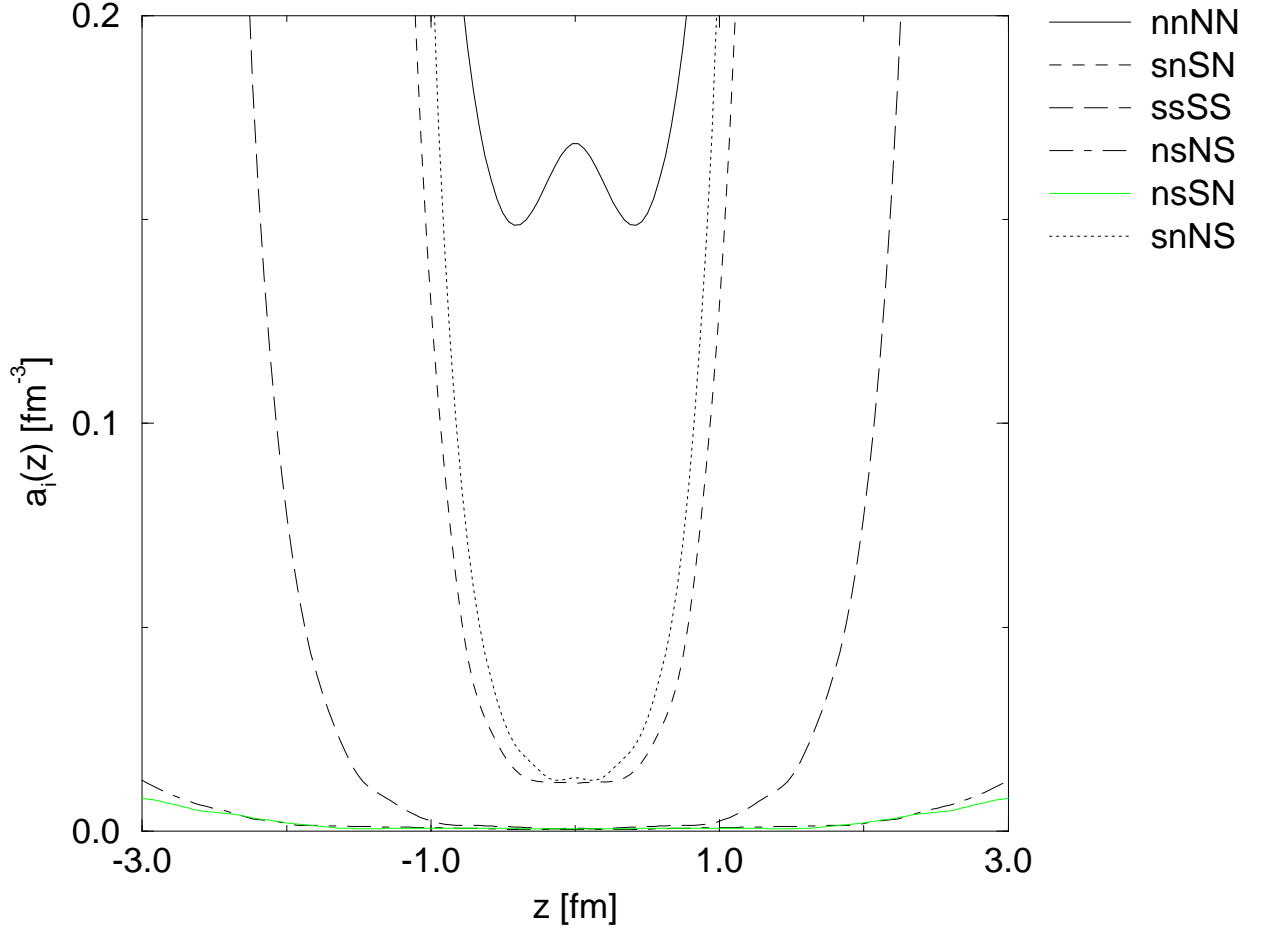}
\caption{
Coefficient functions belonging to the Gaussian
expansion of $\rho(z)$ in (\protect\ref{gaussexp})
for the different flavour components with
$N_{max}=2$.
$n$ denotes an
$u$ or $d$ quark, a capital letter the
corresponding anti-quark.
\label{koeff}}
\end{figure}

At last, we considered the distribution $\rho(z)$ of
a single quark on the $z$-axis given by:
\begin{eqnarray}
\rho(z)=\sum_\alpha\int
d^3\vec x_1 d^3\vec x_2 d^3\vec x_2 d^3\vec x_2\,\,
\left|\Psi_\alpha(\vec x_1, \vec x_2, \vec x_3, \vec x_4)\right|^2
\delta\left(\frac1M\sum_{i=1}^4m_i\vec x_i\right)
\delta(\vec x_1-\vec x)
\end{eqnarray}
where $\vec x_i$ are the absolute coordinates of the
(anti-)quarks.  The first $\delta$-function ensures
that the center of mass is identical with
the center of the coordinate system, and
$\delta(\vec x_1-\vec x)$ cancels the integration
over $\vec x_1$; $\vec x$ is defined by:
\begin{eqnarray}
\vec x = \left(\begin{array}{c}
0 \\ 0 \\ z \end{array}\right).
\end{eqnarray}
Surprisingly, the $\rho(z)$ depends only very
weakly on $N_{max}$, so we can neglect this
dependence for a qualitive analysis.
The distribution has a
quite sharp maximum at $z=0$ and two significant
shoulders at $z=\pm1\mbox{fm}$.
If we make the assumption, that we
have a superposition of two Gaussian-like
quasi-meson distributions with their maxima
seperated by a distance $2a$, we would expect a
distribution of the form
\begin{eqnarray}
e^{-(x+a)^2}+e^{-(x-a)^2}=e^{-x^2-a^2}
\left(e^{+2ax}+e^{-2ax}\right)=ce^{-x^2}\cosh(2ax).
\end{eqnarray}
So we would expect a Gaussian which is modulated by
a cosh-function.  Now it is possible to write
$\rho(z)$ in the form
\begin{eqnarray}
\rho(z)=\sum_ia_i(z)e^{-\frac{z^2}{b_i^2}}
\label{gaussexp}
\end{eqnarray}
where the width $b_i$ is a flavour dependent
constant. The coefficient functions $a_i(z)$
corresponding to the
Gaussian expansion (\ref{gaussexp}) are shown in
the fig.~\ref{koeff}.

It is clear to see, that for $z\rightarrow\pm\infty$
the coefficients which give the important
contributions to the sum behave much like
$\cosh$ functions. This gives evidence that the
$0^0S^{--}$ state in fact describes two free quasi-mesons
and is not a bound $q^2\bar q^2$ system. Although
the contributions of the different flavour
components change, this picture is qualitatively
still correct if one considers the next higher
excited states, also for the $V^\lambda_{\rm
conf}$. That means, {\em all} states could be
most-likely described as free quasi-mesons.

The little local maximum of the $n^2\bar n^2$ coefficient in
fig.~\ref{koeff} may indicate that
there exist a meson-meson attraction for small
distances, which is, however, too weak to lead to a
bound system.

\section{Conclusion}

We have extended a model, which is well
established in the meson and baryon sector,
to $q^2\bar q^2$ systems, without any change in the interaction.

The
interpretation of the results was not easy, because of
the difficulty to distinguish $q^2\bar q^2$ from
bound states and quasi-free mesons. On the basis of
the internal structure, i.e. the
density for a single quark and its representation as
a sum of Gaussians, we believe, however, to have found a good
method to identify possible $q^2\bar q^2$ states in spite of
the background of quasi-free meson states.

On the basis of results presented in this
contribution, the only plausible
interpretation is that {\em all} states we
calculated correspond to quasi-free mesons.
Especially, we found no hint that there exist very tightly
bound states of two quarks and two antiquarks. However, this
does not yet rule out, that the $f_0(975)$ and
$a_0(980)$ mesons are $K\bar K$ molecular states, bound by
long range forces, as provided e.g. by meson exchange
\cite{speth}.

\pagebreak

\pagebreak

\pagebreak

\begin{table}
\caption{
  The parameters of the Hamiltonian, fitted to
  the mass spectra of the light mesons and baryons
  \protect\cite{blask}.}
\hrule
\[
\begin{array}{|c|rc|}
\hline
m_n & 300 & \mbox{Mev}\\
m_s & 540 & \mbox{Mev}\\ \hline
a_{q\bar q} & -892 & \mbox{Mev}\\
b_{q\bar q} & 850 & \mbox{Mev}\, \mbox{fm}^{-1}\\ \hline
a_{qq} & -511 & \mbox{Mev}\\
b_{qq} & 467 & \mbox{Mev}\,\mbox{fm}^{-1}\\ \hline
g & 122 & \mbox{Mev}\, \mbox{fm}^3\\
g'& 82 & \mbox{Mev}\, \mbox{fm}^3\\ \hline
\Lambda& 0.37 & \mbox{fm} \\ \hline
\end{array}
\]
\label{para}
\end{table}

\begin{table}
\caption{The flavour matrix $\tilde g$ with
$g':=\frac{3}{8}g_{\rm eff}(n)$
 and $g:=\frac{3}{8}g_{\rm eff}(s)$. }
\hrule
\[
\begin{array}{c|cc|cccc|ccc|}
 & ud & du & us & ds & sd & su & uu & dd & ss \\ \hline
ud & g & -g & & & & & & & \\
du & -g & g & \multicolumn{4}{c|}{\raisebox{1.5ex}[-1.5ex]{0}} & &
\raisebox{1.5ex}[-1.5ex]{0} & \\ \hline
us & & & g' & 0 & 0 & -g' & & & \\
ds & & & 0 & g' & -g' & 0 & & & \\
sd & \multicolumn{2}{c|}{\raisebox{1.5ex}[-1.5ex]{0}} & 0 & -g' & g' & 0 &
& \raisebox{1.5ex}[-1.5ex]{0} & \\
su & & & -g' & 0 & 0 & g' & & & \\ \hline
uu & & & & & & & & & \\
dd & \multicolumn{2}{c|}{0} & \multicolumn{4}{c|}{0} & & 0 & \\
ss & & & & & & & & & \\ \hline
\end{array}
\]
\label{gqq}
\end{table}

\begin{table}
\caption{The flavour dependent matrix $\hat g$ with $g$ and $g'$
the same as in table~\protect\ref{gqq}.}
\hrule
\[
\begin{array}{c|cc|cccc|ccc|}
 & u\bar d & d\bar u & u\bar s & d\bar s & s\bar d & s\bar u & u\bar u
& d\bar d & s\bar s \\ \hline
u\bar d & -g & 0 & & & & & & & \\
d\bar u & 0 & -g & \multicolumn{4}{c|}{\raisebox{1.5ex}[-1.5ex]{0}} & &
\raisebox{1.5ex}[-1.5ex]{0} & \\ \hline
u\bar s & & & -g' & 0 & 0 & 0 & & & \\
d\bar s & & & 0 & -g' & 0 & 0 & & & \\
s\bar d & \multicolumn{2}{c|}{\raisebox{1.5ex}[-1.5ex]{0}} & 0 & 0 & -g' & 0 &
& \raisebox{1.5ex}[-1.5ex]{0} & \\
s\bar u & & & 0 & 0 & 0 & -g' & & & \\ \hline
u\bar u & & & & & & & 0 & g & g' \\
d\bar d & \multicolumn{2}{c|}{0} & \multicolumn{4}{c|}{0} & g & 0 & g' \\
s\bar s & & & & & & & g' & g' & 0 \\ \hline
\end{array}
\]
\label{gqQ}
\end{table}

\end{document}